# The Interplanetary Magnetic Field and Solar Wind Driven Magnetospheric Reconfiguration

Eugene Savov

*Central Laboratory of Solar-Terrestrial Influences, Bulgarian Academy of Sciences*
*Acad. G. Bonchev Str., Block 3, Sofia 1113, Bulgaria*

eugenesavov@mail.orbitel.bg

**ABSTRACT**

The magnetic disturbances are associated with electric currents as it is well checked at laboratory room scales and described by the Maxwell's equations of electromagnetic field. The analysis of spacecraft observations for more than a quarter of a century failed to provide a self-consistent three-dimensional picture of the solar wind-magnetosphere dynamo generated magnetospheric and ionospheric current systems. The proposed solar wind and the interplanetary magnetic field (IMF) driven reconfiguration of the earth's magnetosphere directly accounts for the observed magnetic disturbances. So role of the magnetospheric currents in creation of the magnetic disturbances is reconsidered in accordance with some poorly understood observations. A quantitative agreement with observations is demonstrated and a laboratory experiment to test the suggested model of the solar wind/IMF-magnetosphere interaction is described.

The solar wind and interplanetary magnetic field (IMF) produced geomagnetic disturbances are believed to originate from electric currents flowing in the earth's ionosphere and magnetosphere. The lack of carries for the field-aligned current system (*Sugiura and Potemra*, 1976; *Hoffman et al.*, 1985) and the ring current (*Roeder et al.*, 1996), the net current between the ionosphere and magnetosphere (*Sugiura and Potemra*, 1976), the uncertain three-dimensional (3D) penetration of the magnetopause boundary (*Siscoe*, 1987; *Lundin*, 1988; *Heikkila*, 1990) and generation and closure of the magnetospheric currents (*Akasofu*, 1984; *Haerendel*, 1990), all these deep problems require introduction of new models for explanation of the ambient space phenomena. The magnetic reconnection (*Dungey*, 1961) is the basic paradigm of modern understanding of energy and momentum transfer into the solar wind/IMF-magnetosphere system (e.g., *Crooker*, 1979; *Kan*, 1988; *Nishida*, 1988, 1989; *Birn et al.*, 1997, 2001). The magnetic reconnection associated 3D magnetospheric circulation between the dayside magnetopause and the magnetospheric tail runs into a topological crisis (*Kennel et al.*, 1989). This confusing situation is mended by smaller scale reconnection processes, called percolation reconnection (*Kennel et al.*, 1989; *Galeev et al.*, 1986). The small-scale processes are unlikely to maintain the self-consistency of the much larger scale one that accounts for the global magnetospheric convection. The threshold strength of the prolonged southward IMF necessary to deepen the magnetic storm (*Russell et al.*, 1974) implies global character of the solar wind/IMF-magnetosphere interaction, which is difficult to discuss in local reconnection terms. The global nature of this interaction is captured in the proposed magnetic tearing, which simply accounts for the puzzling storm-substorm relationship (*Savov*, 1998). Here the magnetic tearing will be discussed as a self-similar and so self-consistent magnetic reconfiguration.

The transverse magnetic disturbances coincide with the visual aurora (e.g., *Kawasaki and Rostoker*, 1979). Large-scale (> 0.5°) transverse magnetic perturbations, considered as region 1 and region 2 field-aligned currents, encircle the geomagnetic pole (*Iijima and Potemra*, 1976). The region 1 current is poleward from the region 2, both currents change direction in each magnetic local time and the dayside of region 1 persists for very low (Kp=0) geomagnetic activity (*Iijima and Potemra*, 1976). The oval pattern of the associated large-scale transverse magnetic disturbance and aurora expands (contracts) after southward (northward) IMF turning (*Bythrow et al.*, 1984; *Nakai et al.*, 1986). The auroral oval expands about 10 times faster after southward IMF turning than it contracts when the IMF becomes northward (*Nakai et al.*, 1986).

The region 1 and region 2 field-aligned currents close in ionospheric areas having different conductivities due to the local time dependence of the solar UV ionisation. Then what keeps the expanding and contracting oval pattern of field-aligned currents stable. The pattern spreads over all local times, suggesting different ionospheric loads for the solar wind-magnetosphere dynamo. What does maintain the oval pattern of these expanding and contracting, having opposite directions and so mutually repelling, region 1 and region 2 field-aligned current sheets? Why is the expansion faster than contraction? The 3D closure and generation of the field-aligned currents is hard to understand. Why does the region 1 current system persist at the dayside during very low geomagnetic activity?

*McDiarmid et al.* (1978) reported "striking similarity" between dawn-dusk transpolar profiles of magnetic disturbances and electric fields. They considered the high latitude tilts of the main field relative to its unperturbed direction. *McDiarmid et al.* (1978) found that the sunward or antisunward direction of the tilt is the same as the direction of the ionospheric convection obtained from the electric field measurements. They explained their findings in terms of magnetic line "foot dragging" in the conducting ionosphere during steady state magnetospheric convection between the dayside magnetopause and the magnetospheric tail. *Sugiura* (1984) discovered very high correlation (0.78 ÷ 0.99) between found to be orthogonal magnetic perturbation field and electric field in the field-aligned current region. The works other authors also indicate closely associated high latitude magnetic disturbances, electric fields and plasma convection (e.g., *Potemra et al.*, 1984).

To understand the solar wind/IMF-magnetosphere interaction we consider first two coupling sources of magnetic field (Fig. 1) and then the solar wind is added (Fig. 2). Figure 1 shows magnetic attraction which corresponds to southward IMF-magnetosphere coupling (*Savov*, 1998). The external field (e.g., the IMF) enters the internal one (e.g., the magnetosphere) through its northern polar region. So during southward IMF (magnetic attraction) the magnetic configuration expands and repels the impinging solar at inner *L*-shells, thus creating the solar wind precipitation and the associated auroral oval at lower latitudes for larger periods of stronger southward IMF. The expansion of the magnetospheric configuration produces the main phase $D_{st}$ negative excursion at middle and low latitudes (*Savov*, 1998). Tailward spreading contractions in the expanded configuration create the near midnight and the near earth substorm expansion onsets (*Savov*, 1998, 2002). The solar wind bends the configuration downstream so creating the magnetotail, whose field lines tilt downstream across the polar cap (Fig. 2). Equatorward from the polar cap the magnetospheric field is likely to bend upstream due to its sheared 3D spiral structure (*Savov*, 2002, p. 127). The upstream tilted field, equatorward from the downstream tilted field of the magnetotail, accounts for the observed oval pattern of large-scale transverse magnetic disturbances, which are usually considered as region 1 and region 2 field-aligned current systems. The large-scale transverse magnetic disturbances, when added to the nearly vertical main field in the polar region, indicate existence of upstream tilt belt. The field-aligned current systems and the enormous difficulties associated with their 3D generation, closure and electric current carriers identification are given a secondary role by the proposed origin of the large-scale transverse magnetic disturbances from magnetic field tilts, which are opposite to the downstream tilted field of the magnetotail. The proposed magnetic reconfiguration accounts for most of the magnetic disturbances leaving a smaller part for magnetospheric currents in accordance with a number of poorly understood observations (e.g., *Sugiura and Potemra,* 1976; *Hoffman et al.*, 1985; *Savov*, 1990; *Roeder et al.*, 1996) that show 10 to 2 times larger magnetic disturbances than what the identified electric current carriers can provide.

The tiltward ionospheric convection, i.e., the **E**×**B** plasma drift, (*McDiarmid et al.*, 1978), the high correlation (0.78 ÷ 0.99) between the found to be orthogonal electric fields and magnetic disturbances in the field-aligned region and also the similarity between the vector plots of the convection and the magnetic disturbances, the convection and the disturbance fields look parallel or antiparallel (*Sugiura*, 1984), all these suggest a common mechanism for generation of the magnetic disturbances, electric fields and plasma drifts. The configuration of the magnetosphere in the absence of the solar wind and the IMF forcings is called unperturbed. The shape of the magnetosphere is created from solar wind and IMF generated reconfiguration lines that continuously tug the main field lines from their unperturbed direction. The reconfiguration lines generate electric fields and drive the ionospheric convection tiltward (*Savov*, 2002, p. 139), thus creating the closely associated magnetic disturbances, electric fields and ionospheric convection.

Figures 1 and 2 show the solar wind and IMF driven magnetospheric reconfiguration, which simply produces the transverse magnetic disturbances. The main field self-similarly and so self-consistently reconfigures due to the applied stresses (the solar wind and the IMF). In this way it tilts from its unperturbed direction and so creates the oval pattern of large-scale transverse magnetic disturbances and auroral precipitation. The southward (northward) IMF and the earth's magnetosphere couple like two attracting (repelling) sources of magnetic field (*Savov*, 1998), so creating the observed expansion (contraction) of the auroral oval and the associated oval pattern of large-scale transverse magnetic disturbances, which are likely to originate from upstream main field tilts, appearing equatorward from the downstream tilted field of the magnetotail. The ionospheric roots of the magnetotail field lines are tilted downstream and so the magnetospheric field lines around those of the tail will tilt upstream, in accordance with the obtained from *Iijima and Potemra* (1976) oval pattern of large-scale transverse magnetic disturbances, surrounding the northern geomagnetic pole at ionospheric altitudes. The spacecraft obtained large-scale transverse magnetic disturbances (e.g., Fig. 4 of *Zanetti et al.*, 1984), when added to the nearly vertical undisturbed field of the polar region, indicate downstream tilts across the both polar caps (the areas in which the northern and southern parts (lobes) of the magnetotail are rooted) and opposite (upstream) tilts around them. It is shown that the ionosphere convects tiltward in accordance with the expansion and contraction of the magnetosphere, and the northward IMF produced contraction (Fig. 3) accounts for the observed high latitude near noon sunward convection, due to the sunward main field tilt in the local noon sector of the contracted configuration (*Savov*, 2002, p. 147).

The IMF enters in the northern dayside cusp region and expands the magnetic configuration as shown in Figure 1. The puzzling findings that magnetic substorms weaken rather than enhance magnetic storm (*Iyemori and Rao*, 1996) and the main phase $D_{st}$ decrease "well before" the first main phase substorm expansion onset (*McPherron*, 1997) are directly explained with the IMF driven expansion and contraction of the magnetic configuration (*Savov*, 1998). The magnetic substorm expansion is viewed as an outward (tailward) spreading contraction in the southward IMF expanded configuration. This explains the poorly understood near earth source of the substorm expansion onset, indicated by the onset initiation with the brightening of the most equatorward auroral arc. The absence of magnetic activities in the magnetotail before the auroral onset suggests substorm activity beginning first near the Earth and later in the mid-tail (*Lui et al.*, 2000).

The persistence of the expanded configuration will account for the observed longer time of auroral oval contraction after northward IMF turning than its expansion after southward IMF reversal. The direct southward (the indirect northward) IMF entry through the northern dayside cusp region will create faster expansion (slower contraction) of the magnetic configuration and the auroral oval. The multiscale expansions and contractions of the magnetosphere create magnetic storms and substorms (*Savov*, 1998). The proposed model of the solar wind/IMF – magnetosphere interaction can be experimentally tested as described in Fig. 4. The model predicts that the field, which simulates the IMF, will expand (contract) the simulated magnetosphere and artificial auroral oval during attraction (repulsion) between the two magnetic coils (Fig.4). The expansion will be faster than the contraction.

The average distance to the dayside magnetopause is about 10 Earth radii (10 $R_E$). The IMF enters the magnetosphere through its northern dayside cusp region (*Savov*, 1998). Then the entering southward IMF is nearly perpendicular to a surface encircled by a contour a having radius of about 10 $R_E$. The flux of the southward IMF $B_z$=−5 nT component through this surface is five times larger than the magnetic flux produced from field having strength 100 nT through a contour having a radius of 1 $R_E$.

The latter magnetic flux will create the global low latitude depression of the geomagnetic field, corresponding to a magnetic storm of magnitude $D_{st}=-100$ nT. Hence the southward IMF flux entry could expand the magnetospheric configuration as shown in Fig. 1 and thus to account directly for the storm-time $D_{st}$ negative excursion of the geomagnetic field.

The quantitative assessment of the magnetic storm associated southward IMF flux entry into the magnetosphere is in agreement with observations, thus confirming the proposed magnetic reconfiguration. The magnetosphere expands (contracts) for southward (northward) IMF so accounting for the expansion (contraction) of the auroral oval and the coincident oval pattern of large-scale transverse magnetic disturbances and also for the creation of magnetic storms and substorms (Figs. 1-3). The electrically charged particles enter the disturbed regions of the magnetosphere and so create the field-aligned and ring currents. Hence the magnetospheric currents and the magnetic disturbances they generate turn to be secondary to the proposed magnetic reconfiguration created magnetic disturbances (Figs. 1,2). Then the magnetic disturbances generated from these currents are likely to be smaller than their source the initial magnetic disturbances created from the solar wind and the IMF driven global magnetic reconfiguration (Figs. 1,2). In this way the poorly understood and so usually neglected observations of lack of current carriers for the field-aligned and the ring currents (*Sugiura and Potemra*, 1976; *Hoffman et al.*, 1985; *Savov*, 1990; *Roeder et al.*, 1996) are explained.

The unusually long growth phase of an isolated magnetic substorm, preceded by a period of a steady northward IMF (*Lui et al.*, 1998), is explained with northward IMF produced contraction of the magnetospheric configuration (Fig. 3). It will take more time for the southward IMF to expand (Figs. 1 and 2) the more contracted configuration, created by a long period of steady northward IMF. This explains the observed puzzling longer growth phase of the isolated substorm. The southward IMF expands the magnetosphere, thus creating the growth phase of the substorm, then a partial contraction in the magnetotail generates the substorm expansion. The longer period of southward IMF generates the main phase of magnetic storm (*Kamide et al.*, 1977; *Gonzalez* and *Tsurutani*, 1987) on which the strongest substorms will be imposed as greater partial tailward spreading contractions in the expanding and so less stable configuration (*Savov*, 1998).

The solar wind and the IMF driven magnetospheric reconfiguration (Figs. 1,2) accounts directly for the general picture of the solar wind/IMF-magnetosphere interaction and some poorly understood aspects of magnetic storm-substorm relationship and confusing observations. The generated magnetic reconfiguration lines create the observed closely associated magnetic disturbances, electric fields, plasma convection, precipitation and upflows (*Savov*, 1998, 2002). The proposed magnetic reconfiguration (Figures 1-3) offers deeper insights in the structure of magnetic interaction and the solar-terrestrial relationship. The obtained new understanding is developed in the theory of interaction (*Savov*, 2002), which goes far beyond the current knowledge of the geospace and the macro and micro universe.

7

**FIGURE CAPTIONS**

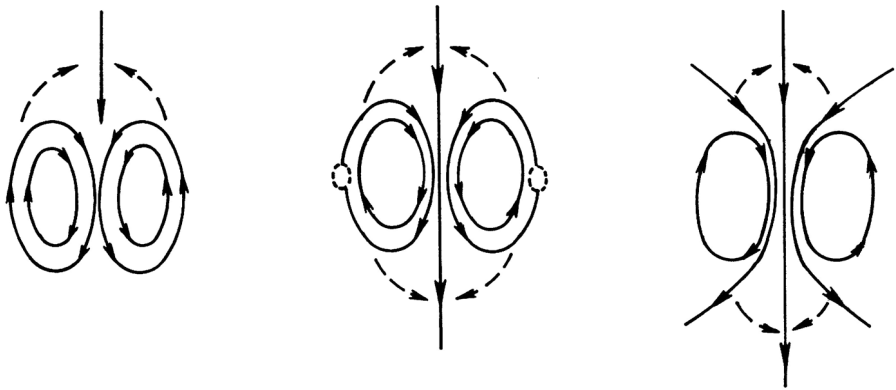

**Fig. 1.** The magnetic lines of force enter the northern magnetic pole and exit from the southern pole of every magnetic configuration (the thick line arrows). Two sources of magnetic field meet their opposite poles. The external field, e.g. the southward IMF, enters directly the magnetic configuration through its northern polar region. Then the configuration expands as its reconfiguration lines move poleward (the dashed arrows) and tear near the magnetic equator (the dashed curve). The field of the expanded configuration strengthens poleward and weakens equatorward, thus accounting for the general picture of the solar wing/IMF-magnetosphere interaction and the puzzling magnetic storm-substorm relationship. The solar wind is repelled at inner (outer) lines in the expanded (contracted)

configuration, thus creating expanding (contracting) auroral oval, larger (smaller) magnetotail and D$_{st}$ decrease (increase) and also high (low) latitude increase (decrease) of the strength of the vertical geomagnetic field component near the auroral region. The magnetic configuration reconfigures self-similarly and so self-consistently by expansions and contractions, which depending on their scales create magnetic storms and substorms.

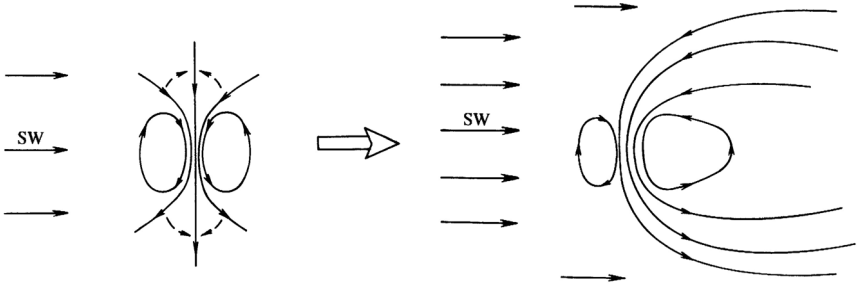

**Fig. 2.** The impinging solar wind (SW) creates the downstream tilted lines of the magnetotail and the upstream tilt belt equatorward from them, consisting of upstream tilted main field lines. The solar wind particles are repelled at inner lines in the more expanded configuration and so the magnetotail becomes larger and the boundaries of the auroral precipitation shift equatorward. The upstream tilt belt and the auroral oval move equatorward with the expansion of the magnetosphere during southward IMF because then the magnetotail becomes larger (Fig. 1). The increase of the magnetospheric expansion during a longer period of stronger southward IMF (Fig. 1) will produce directly the beginning of the magnetic storm main phase D$_{st}$ decrease before the first main phase substorm expansion onset. Later, in agreement with the poorly understood observations of *Iyemori and Rao* (1996), substorms as tailward spreading contractions in the magnetosphere, will decrease the global magnetospheric expansion and thus weaken the magnetic storm. The decrease of magnetic configuration expansion after northward IMF turning will lead to upstream tilt belt and auroral oval contraction, and also to the observed D$_{st}$ increase during the recovery phase of the magnetic storm.

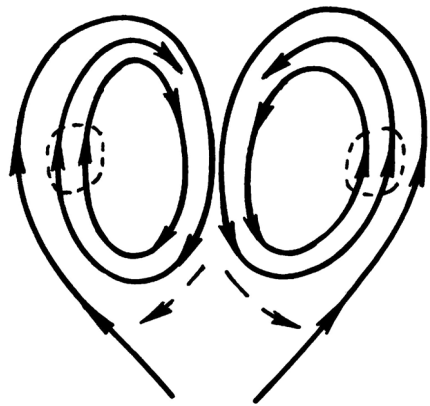

**Fig. 3.** The external field, e.g., the northward IMF, enters indirectly through the northern polar region of the magnetic configuration and winds up 3D spirally into it. The configuration contracts, thus repelling the solar wind at outer lines (the dashed curve), so creating the observed greater distance to the dayside magnetopause and the contraction of the auroral oval after northward IMF turning. More The configuration becomes more contracted during a longer period of a stronger northward IMF. Afterwards it will take more time for the southward IMF to expand this configuration and so a longer growth phase for an isolated substorm will be produced in agreement with the puzzling observations of *Lui et al.* (1998).

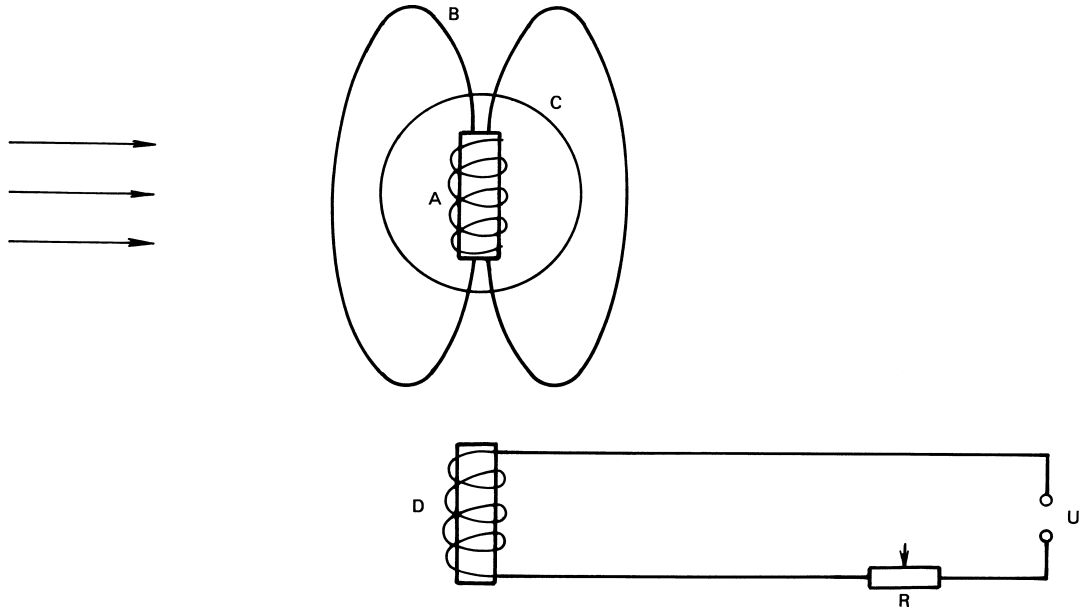

**Figure. 4.** Experimental test for the proposed magnetospheric reconfiguration to the IMF and the impinging solar wind**.** Electrons shown by the three arrows impinge on the magnetic configuration B), generated by the magnetic coil A), which is inserted into the sphere C) that is coated with a fluorescent paint. The magnetic poles of the field created from the coil A) are put on one straight line with the magnetic poles of the field generated from the secondary coil D), which is connected to a circuit containing rheostat R) and the direct current source U). The velocity of the electrons is parallel to the equatorial plane of the magnetic field configuration B), produced by the coil A). The field from the magnetic coil D) simulates the interplanetary magnetic field, whose crucial importance for the geomagnetic phenomena was not known when the famous Birkeland terrella experiment (e.g., *Egeland*, 1984) was performed at the end of nineteenth century. The suggested magnetic reconfiguration (Figs. 1-3) predicts that the obtained artificial auroral oval should expand (contract) when the magnetic configurations generated by the two coils meet their opposite (same) magnetic poles. (from *Savov*, 2002, p. 195)